\documentclass[fleqn,usenatbib]{mnras}

\usepackage{newtxtext,newtxmath}
\usepackage[T1]{fontenc}
\usepackage{ae,aecompl}

\usepackage{graphicx}
\usepackage{amsmath}
\usepackage{bm}
\usepackage{comment}
\usepackage[dvipsnames]{xcolor}

\newcommand{\Msun}{M_{\sun}}
\newcommand{\vir}{\mathrm{vir}}
\newcommand{\de}{\delta_\mathrm{e}}

\title[BP Algorithm]
{Baryon Pasting Algorithm: Halo-based and Particle-based Pasting Methods}

\author[Osato \& Nagai]{
Ken Osato$^{1,2,3,4}$\thanks{E-mail: ken.osato@chiba-u.jp}
and Daisuke Nagai$^{5}$
\\
$^{1}$Center for Frontier Science, Chiba University,
1-33 Yayoi-cho, Inage-ku, Chiba 263-8522, Japan\\
$^{2}$Department of Physics, Graduate School of Science, Chiba University,
1-33 Yayoi-cho, Inage-ku, Chiba 263-8522, Japan\\
$^{3}$Center for Gravitational Physics and Quantum Information,
Yukawa Institute for Theoretical Physics, Kyoto University,\\
Kitashirakawa Oiwakecho, Sakyo-ku, Kyoto 606-8502, Japan\\
$^{4}$LPENS, D\'epartement de Physique, \'Ecole Normale Sup\'erieure,
Universit\'e PSL, CNRS, Sorbonne Universit\'e, Universit\'e de Paris,\\
24 rue Lhomond, 75005 Paris, France\\
$^{5}$Department of Physics, Yale University, New Haven, CT 06520, USA\\
}

\date{Accepted XXX. Received YYY; in original form ZZZ\\
Report number: YITP-21-156}

\pubyear{2022}

\begin{document}
\label{firstpage}
\pagerange{\pageref{firstpage}--\pageref{lastpage}}
\maketitle

\begin{abstract}
We present a fast methodology to produce mock observations
of the thermal and kinetic Sunyaev--Zel'dovich (SZ) effects
based on the dark matter only $N$-body simulations
coupled with the analytic intra-cluster medium model.
The methods employ two different approaches: halo-based pasting (HP)
and particle-based pasting (PP).
The former pastes gas density and pressure onto halos and requires only a halo catalogue,
and the latter considers the contribution from field particles as well,
i.e., particles that do not belong to any halos and thus utilise the full particle information.
Therefore, the PP algorithm incorporates secondary effects beyond the HP algorithm:
asphericity of halos and contribution from diffuse gas.
In particular, such a diffuse component is the dominant source of the kinetic SZ effect.
As validation of our methods, we have produced 108 all-sky maps with HP
and 108 flat-sky maps, which cover $5 \times 5 \, \mathrm{deg}^2$ with both HP and PP,
and measured power spectra of the maps.
Our method can produce a mock map within a few hours, even for all-sky coverage with a parallel computational environment.
The power spectra of HP maps are consistent with the halo model prediction of the thermal SZ effect.
On the other hand, the power spectra of PP maps are suppressed due to the halo asphericity but
can reproduce better the theoretical prediction for the kinetic SZ effect.
We discuss the utility of baryon-pasted mock SZ maps
for estimating the covariance matrix of SZ statistics and modelling the selection
and projection effects for cluster cosmology.
\end{abstract}



\section{Introduction}
\label{sec:introduction}
The large-scale structures of the Universe contain rich information
about the physics of structure formation driven by the gravity of the dark matter
and accelerated expansion of dark energy.
In the coming decades, a plethora of multi-wavelength cosmological surveys will be ongoing and
underway to study cosmology and the structure formation of the Universe.
The Sunyaev--Zel'dovich (SZ) effects \citep{Sunyaev1970,Sunyaev1972,Sunyaev1980} have been emerging
as promising probes of cosmology and astrophysics.
The SZ effects leave imprints in temperature and polarisation variation
of cosmic microwave background (CMB) through inverse Compton scattering
between CMB photons and free electrons.
There are two classes for the SZ effects \citep[][for reviews]{Birkinshaw1999,Carlstrom2002,Kitayama2014,Mroczkowski2019}.
The thermal SZ (tSZ) effect is induced by hot electrons in galaxy clusters,
making it a powerful method for finding galaxy clusters out to high redshift.
The (late-time) kinetic SZ (kSZ) effect, on the other hand,
is the temperature variation in CMB produced by the peculiar velocity of free electrons
with respect to the CMB frame.
The amplitude of the kSZ signal is lower than the tSZ effect,
but it contains information about the cosmic velocity field.

Observationally, the tSZ effect has been detected
for several hundreds and thousands of galaxy clusters
by space and ground-based telescopes, such as
\textit{Planck} \citep{PlanckCollaborationXXII,PlanckCollaborationXXIV},
Atacama Cosmology Telescope \citep[ACT;][]{Madhavacheril2020,Hilton2021},
and South Pole Telescope \citep[SPT;][]{Bleem2015,Bocquet2019,Huang2020,Bleem2020}.
Measurements of the kSZ effect are more challenging compared to the tSZ effect
because the kSZ effect is comparatively much smaller, and its signal does not depend on frequency.
However, recent development in novel techniques enabled the detection of
kSZ effects \citep{Hand2012,Ade2016,Soergel2016,DeBernardis2017},
whose detection significance has been improving rapidly.

Since the SZ effects are sourced by free electrons, modelling of the SZ effects requires
cosmic gas density and pressure distributions.
For the hot gas in virialized dark matter halos,
modern hydrodynamical cosmological simulations can capture the impacts of galaxy formation
on observable properties of massive dark matter halos \citep{Springel2001a,Dolag2005,
Nagai2006,Nagai2007,Battaglia2012a,Battaglia2012b,McCarthy2014,Dolag2016,Soergel2018,Coulton2022}.
Physical insights from these simulations have enabled the development of
an ever more physically-motivated and computationally efficient model of
the intra-cluster medium (ICM) \citep{Komatsu2001,Ostriker2005,Bode2007,Bode2009,Shaw2010,Schneider2019,Arico2020}
that can be used to create mock observations of multi-wavelength cosmological surveys.

Fast production of mock SZ observations is a powerful tool
for SZ measurements \citep[e.g.,][]{Sehgal2010,Flender2016,Stein2020}.
Recently, novel approaches based on machine learning
have been proposed \citep{Troester2019,Thiele2020,Han2021,Rothschild2022}.
A large number of independent realisations enables us to
estimate the covariance matrix of SZ statistics
\citep[for an analytical approach, see][]{Osato2021}.
We can apply a survey mask and add detector noise
by post-processing the mock maps and quantifying various systematic uncertainties,
such as the selection function and projection effects.
To realise fast production of mock SZ simulations,
we propose a novel method to simulate the gas distribution
based on a physically-motivated, computationally efficient
analytic ICM model \citep{Ostriker2005,Bode2007,Bode2009,Shaw2010}
combined with dark matter only $N$-body simulations,
which can be run much faster than hydrodynamical simulations.
This ICM model also adopts halo model prescription \citep[see][for a review]{Cooray2002},
where all matter and gas are associated with halos.
This assumption holds for observables sourced by hot gas, e.g., the tSZ effect.
On the other hand, some observables, e.g., the kSZ effect, are sensitive to the diffuse gas component.
In this paper, we focus on this particular ICM model, but our pasting methods
are compatible with other analytic ICM models or fitting formulas
of thermodynamic profiles of ICM.
Therefore, our methods have the potential to generate a suite of mock simulations
with various ICM models.

In this paper, we present two different approaches to paste gas distribution
onto dark matter only $N$-body simulations:
halo-based pasting (HP) and particle-based pasting (PP).
The former follows the halo model prescription;
all gas resides in halos, and only the halo catalogue is required to produce mock simulations.
In addition to halos, the latter considers all particles in simulations,
including the \textit{field} component lying outside of halos.
This PP method also captures the aspherical gas distribution within halos,
which is missing in previous pasting methods but has appreciable impacts
on SZ effects.
Furthermore, though in the HP method, the velocity of particles is assumed to be uniform within halos
and follow the mean bulk velocity, the kinematics within halos has been considered in the PP method.
These secondary effects on the SZ effects,
which are missing in most previous works, are properly considered in our PP method.
For example, in \citet{Flender2016}, the contribution from the field component has been incorporated, but for particles that belong to halos, the gas is painted based on spherically symmetric halos.
The prototypes of these methods have been applied in precedent works \citep{Osato2018,Osato2020}.
To assess the performance of these two methods,
we create 108 full-sky and patch ($5 \times 5 \, \mathrm{deg}^2$)
mock maps of SZ effects
based on halo-based and particle-based pasting methods.
To verify these methods, we measure the power spectra of mock maps
and compare them with the theoretical predictions.
Although this paper focuses on SZ effects, our methods can easily be extended to other observables,
such as X-ray emission and dispersion measures probed by fast radio bursts,
and cross-correlations between these observables.

This paper is organised as follows.
In Section~\ref{sec:model}, we provide a brief overview of the analytical modelling of ICM.
In Section~\ref{sec:map_making}, we present simulations and the map-making algorithms.
In Section~\ref{sec:theory}, we review theoretical models of tSZ and kSZ power spectra.
Section~\ref{sec:results} discusses the resultant mock maps and their power spectra.
We conclude in Section~\ref{sec:conclusions}.

Throughout this paper, we adopt the flat $\Lambda$ cold dark matter Universe
and cosmological parameters inferred from CMB temperature and polarisation anisotropies of WMAP 9 years results \citep{Hinshaw2013}:
baryon density $\Omega_\mathrm{b} = 0.046$,
matter (cold dark matter + baryon) density $\Omega_\mathrm{m} = 0.279$,
Hubble parameter $H_0 = 70 \, \mathrm{km} \, \mathrm{s}^{-1} \, \mathrm{Mpc}^{-1}$,
the amplitude of matter fluctuation at $8 \, h^{-1} \, \mathrm{Mpc}$ $\sigma_8 = 0.82$,
and the slope of the scalar perturbation $n_\mathrm{s} = 0.97$.

\section{Model of Intra-Cluster Medium}
\label{sec:model}
We present the analytic model of thermodynamic profiles of ICM within a halo.
The key assumption of this model is that
the polytropic gas is in hydrostatic equilibrium in the dark matter halos.
The amplitudes of thermodynamic profiles are determined with the three boundary conditions:
energy conservation, pressure boundary condition, and gas mass conservation.
The non-gravitational effects, such as feedback
from supernovae and supermassive black holes \citep{Ostriker2005,Bode2007,Bode2009}
and the non-thermal pressure by turbulent gas motions \citep{Shaw2010},
contribute to the total energy budget.
The latest analytic ICM model has also been calibrated
using Chandra observations of SZ selected clusters \citep{Flender2017}.

\subsection{Profiles of thermodynamic quantities}
In this model, we adopt a spherically symmetric dark matter halo with the total mass distribution
given by Navarro--Frenk--White (NFW) profile
\citep{Navarro1996,Navarro1997}:
\begin{equation}
\label{eq:NFW_profile}
  \rho_\mathrm{tot} (r) = \frac{\rho_\mathrm{s}}{(r/r_\mathrm{s}) (1 + r/r_\mathrm{s} )^2},
\end{equation}
where $\rho_\mathrm{s}$ and $r_\mathrm{s}$ are scale density and radius,
respectively.\footnote{In previous literature, the total density profile
is denoted as a ``dark matter'' density profile.
Here, we consider total matter (dark matter, gas, and stars) density profile,
and thus we use the notation of ``total'' density profile instead.}
We adopt the virial radius $r_\vir$ as the halo radius,
and the virial radius is determined as
\begin{equation}
  M_\vir = \frac{4 \pi}{3} r_\vir^3 \Delta_\vir (z) \rho_\mathrm{cr} (z),
\end{equation}
where the virial overdensity $\Delta_\vir (z)$ is given in \citet{Bryan1998},
and $\rho_\mathrm{cr} (z)$ is the critical density.
We truncate the density profile at the virial radius.
The scale radius is known to correlate strongly with the halo mass,
and the scale radius can be determined through the concentration parameter:
\begin{equation}
\label{eq:concentration}
  c_\mathrm{vir} \equiv \frac{r_\vir}{r_\mathrm{s}} ,
\end{equation}
where we adopt the fitting formula calibrated with $N$-body simulations in \citet{Klypin2016}.
Then, the scale density is given by
\begin{equation}
  \rho_\mathrm{s} = \frac{M_\vir}{4 \pi r_\mathrm{s}^3 m (c_\vir)} ,
\end{equation}
where
\begin{equation}
  m(c) \equiv \int_0^c \frac{x}{(1+x)^2} \mathrm{d}x = \ln (1+c) - \frac{c}{1+c}.
\end{equation}
Thus, given the virial mass, the density profile is specified.

Next, we derive profiles of the gas component.
The basic equation for gas profile is the Euler equation:
\begin{equation}
  \frac{\mathrm{d} P_\mathrm{tot} (r)}{\mathrm{d} r} = -\rho_\mathrm{g} (r) \frac{\mathrm{d} \Phi (r)}{\mathrm{d} r},
\end{equation}
where $P_\mathrm{tot} (r)$ is the total gas pressure, $\rho_\mathrm{g} (r)$ is the gas density,
and $\Phi (r)$ is the gravitational potential.
We assume that the gravitational potential is sourced by the total matter density
and the explicit expression for the NFW profile is derived
in Eq.~(9) of \citet{Lokas2001}.\footnote{Note that in \citet{Lokas2001},
the density profile is not truncated at the virial radius.
For the truncated profile, a constant must be added to the potential
to ensure the potential vanishes at infinity.}
The key assumption in this model is that gas follows the polytropic relation:
\begin{equation}
  P_\mathrm{tot} (r) = P_0 \left( \frac{\rho_\mathrm{g} (r)}{\rho_0} \right)^\Gamma ,
\end{equation}
where $\Gamma$ is the adiabatic index and we fix $\Gamma = 1.2$ \citep{Shaw2010},
and $\rho_0$ and $P_0$ are the density and pressure at the centre, respectively.
By solving the Euler equation, the pressure and gas density profiles are given by
\begin{align}
  P_\mathrm{tot} (r) &= P_0 \theta^{n+1} (r) , \\
  \rho_\mathrm{g} (r) &= \rho_0 \theta^{n} (r) ,
\end{align}
where $n = 1/\Gamma - 1$ is the polytropic index.
The polytropic function $\theta (r)$ is given by
\begin{equation}
  \theta (r) = 1 + \frac{\Gamma - 1}{\Gamma} \frac{\rho_0}{P_0}
  [\Phi_0 - \Phi (r)] ,
\end{equation}
where $\Phi_0$ is the gravitational potential at the centre.

\subsection{Boundary conditions}
To determine the free parameters, we impose the following boundary conditions:
\begin{itemize}
  \item energy conservation
  \item pressure boundary condition
  \item gas mass conservation
\end{itemize}
First, let us consider energy conservation through the ``rearrangement'' of gas;
gas initially follows the total matter density profile (Eq.~\ref{eq:NFW_profile})
and then, the gas distribution is redistributed due to various physical processes.
In addition to dark matter and gas, a halo has a stellar component.
The stellar mass is derived from the stellar-to-halo-mass relation $F_* (M_{500}, z)$,
which is parameterised as
\begin{equation}
  F_* (M_{500}, z) \equiv \frac{M_* (<r_{500})}{M_{500}}
  = f_* \left( \frac{M_{500}}{3 \times 10^{14} \Msun}\right)^{-S_*} ,
\end{equation}
where $r_{500}$ is the radius within which the mean density is 500 times the critical density,
$M_{500} = (4 \pi /3) 500 \rho_\mathrm{cr} r_{500}^3$ is the corresponding mass,
$f_*$ is the stellar mass fraction at the pivot mass $3 \times 10^{14} \Msun$,
$S_*$ is the slope of mass, and we assume there is no redshift dependence for the relation.
This relation is applicable within $r_{500}$, but we assume
the ratio is the same as the one at the virial radius, since a large fraction of mass
and stellar mass is already within $r_{500}$, and extrapolating up to the virial radius has a minor impact.
Thus, the total stellar mass is given by $M_* = F_* (M_{500}, z) M_\mathrm{vir}$.
The total energy is the sum of the kinetic and gravitational potential energy of the matter
embedded in the NFW profile:
\begin{equation}
  \label{eq:initial_energy}
  E_{\mathrm{g}, i} = f_\mathrm{b} \left[ 2 \pi \int_{r_*}^{r_\vir} \!\!
  \rho_\mathrm{tot} (r) 3 \sigma^2 (r) r^2 \mathrm{d}r + \int_{r_*}^{r_\vir} \!\!
  \Phi (r) \frac{\mathrm{d}M}{\mathrm{d}r} \mathrm{d}r \right] ,
\end{equation}
where $f_\mathrm{b} = \Omega_\mathrm{b}/\Omega_\mathrm{m}$ is
the universal baryon fraction,
and $\sigma^2 (r)$ is the velocity dispersion profile
(for an analytic expression, see Eq.~13 of \citet{Lokas2001} in an isotropic orbit).
Before rearrangement, the stars are assumed to form within the stellar radius $r_*$,
which is determined through the relation $f_\mathrm{b} M_\mathrm{tot} (< r_*) = M_*$,
and the total enclosed mass $M_\mathrm{tot} (< r)$
is given by
\begin{equation}
  M_\mathrm{tot} (< r) = \int_0^r \!\! 4 \pi r^2 \rho_\mathrm{tot} (r) \mathrm{d}r .
\end{equation}
The gas component receives energy from the dark matter component through
dynamical friction and from the stellar component through feedback processes
such as supernova explosions.
As a result of the rearrangement,
the gas profile has a definite boundary,
which is denoted as $r_f$.
In general, $r_f$ is larger than the halo virial radius $r_\vir$
because gas expands due to energy injection from feedback processes.
Such gas expansion effect has been considered in
halo model approaches to model baryonic effects \citep[see, e.g.,][]{Schneider2015}.
Energy conservation is given by
\begin{equation}
  \label{eq:energy_conservation}
  E_{\mathrm{g}, i} = E_{\mathrm{g}, f} + \epsilon_\mathrm{DM} |E_\mathrm{DM}|
  + \epsilon_* M_* c^2 + \Delta E_p,
\end{equation}
where the left-hand side is the initial energy,
and the right-hand side is the sum of the kinetic and potential energy of the gas
after rearrangement $E_{\mathrm{g}, f}$,
injection from the dark matter component
$\epsilon_\mathrm{DM} |E_\mathrm{DM}|$,
and the work due to rearrangement $\Delta E_p$.
The energy of gas after rearrangement is given by
\begin{equation}
  E_{\mathrm{g}, f} = \int_0^{r_f} \!\! 4 \pi
  \left[ \frac{3}{2} P_\mathrm{tot} (r) +\rho_\mathrm{g} (r) \Phi(r) \right] r^2 \mathrm{d}r .
\end{equation}
The energy of dark matter component $E_\mathrm{DM}$
is calculated by substituting $r_* \to 0$ and $f_\mathrm{b} \to 1$
in Eq.~\eqref{eq:initial_energy},\footnote{Exactly speaking, the energy with $r_* \to 0$ and $f_\mathrm{b} \to 1$
corresponds to ``the energy if all matter is dark matter''.
In reality, the dark matter fraction is $f_\mathrm{DM} = 1 - f_\mathrm{b}$
and the energy of the dark matter component corresponds to the one with $r_* \to 0$ and $f_\mathrm{b} \to f_\mathrm{DM}$.
However, the mismatch can be absorbed into the amplitude parameter $\epsilon_\mathrm{DM}$ and
we keep using this definition of $E_\mathrm{DM}$ in \citet{Shaw2010}.}
and $\epsilon_\mathrm{DM}$ and $\epsilon_*$ are free parameters that control
energy transfer from dark matter and stellar components, respectively.
The pressure at the boundary should be the surface pressure $P_s$:
\begin{equation}
  \label{eq:surface_pressure}
  P_s = f_\mathrm{b} \rho_\mathrm{tot} (r_\vir) \sigma^2 (r_\vir) = P_\mathrm{tot} (r_f) .
\end{equation}
Finally, the work done by the surface pressure is given by
\begin{equation}
  \Delta E_p = \frac{4 \pi}{3} \left( r_\vir^3 - r_f^3 \right) P_s .
\end{equation}
The gas mass should be conserved within $r_f$:
\begin{equation}
  \label{eq:gas_mass}
  f_\mathrm{b} M_\vir = \int_0^{r_f} 4 \pi r^2 \rho_\mathrm{g} (r) \mathrm{d} r + M_*.
\end{equation}
Thus, there are three parameters ($\rho_0, P_0, r_f$) with three constraints
(Eqs.~\ref{eq:energy_conservation}, \ref{eq:surface_pressure}, and \ref{eq:gas_mass}).
These parameters are solved to satisfy the above three relations
with the Newton--Raphson method.

\subsection{Non-thermal pressure}
A considerable fraction of pressure is caused by
non-thermal processes, e.g., magnetic field, cosmic rays, or turbulence \citep{Nelson2014}.
For the tSZ effect, only thermal pressure contributes to the signal.
We incorporate this effect with the simple fitting formula of the non-thermal pressure profile:
\begin{equation}
  R_\mathrm{nt} (r, z) \equiv \frac{P_\mathrm{nt}}{P_\mathrm{tot}} = \alpha_\mathrm{nt} f(z)
  \left( \frac{r}{r_{500}} \right)^{n_\mathrm{nt}},
\end{equation}
where the amplitude function $f (z)$ is given by
\begin{equation}
  f (z) = \mathrm{min} \left[ (1+z)^{\beta_\mathrm{nt}},
  (f_\mathrm{max}-1) \tanh (\beta_\mathrm{nt} z) + 1 \right] ,
\end{equation}
where we set $f_\mathrm{max} = 4^{-n_\mathrm{nt}} / \alpha_\mathrm{nt}$
following \citet{Shaw2010}.
Then, the thermal pressure $P_\mathrm{th}$ can be given by
\begin{equation}
  P_\mathrm{th} (r, z) = P_\mathrm{tot} (r, z) \times \mathrm{max}
  \left[ 0, 1 - R_\mathrm{nt}(r, z) \right] .
\end{equation}

In this model, we have introduced free parameters, which must be calibrated
against observations and simulations.
We adopt parameters calibrated in \citet{Shaw2010} and \citet{Flender2017}.
In Table~\ref{tab:ICM_params}, we summarise the parameters used in this model.

\begin{table}
  \centering
  \caption{
    The summary of parameters of the ICM model.
    These parameters are calibrated with observations in \citet{Shaw2010,Flender2017}.
  }
  \label{tab:ICM_params}
  \begin{tabular}{ccl}
    \hline
    Symbol & Value & Definition \\
    \hline
    $\Gamma$ & $1.2$ & Adiabatic index \\
    $\epsilon_\mathrm{DM}$ & $0$ & Energy transfer between gas and dark matter \\
    $\epsilon_*$ & $3.97 \times 10^{-6}$ & Stellar feedback parameter \\
    $f_*$ & $0.026$ & Amplitude of stellar mass fraction \\
    $S_*$ & $0.12$ & Slope of stellar mass fraction \\
    $\alpha_\mathrm{nt}$ & $0.18$ & Amplitude of non-thermal pressure \\
    $\beta_\mathrm{nt}$ & $0.5$ & Redshift dependence of non-thermal pressure \\
    $n_\mathrm{nt}$ & $0.8$ & Radial slope of non-thermal pressure \\
    \hline
  \end{tabular}
\end{table}

\section{Map-Making Algorithm}
\label{sec:map_making}
To generate mock tSZ and kSZ maps, we post-process halo catalogues or particle snapshots
generated from $N$-body simulations.
We present two different algorithms: halo-based pasting (HP) and particle-based pasting (PP).
The former pastes gas onto halos and requires only halo catalogues,
while the latter pastes gas onto all particles within and outside of halos.

\subsection{Simulation}
\label{sec:simulation}
For HP, we use all-sky halo catalogues produced in \citet{Takahashi2017}.
The $N$-body simulations are run with \texttt{L-Gadget2} code \citep{Springel2005}
and halos are identified with \texttt{Rockstar} algorithm \citep{Behroozi2013}.
These catalogues contain halos up to the redshift $z = 3.65$ and
there are 108 pseudo-independent realisations.
We consider only halos whose virial mass is larger than $5 \times 10^{12} \, h^{-1} \, \Msun$
because the ICM model is not calibrated against low-mass halos.\footnote{Though the contributions
from such low-mass halos to auto-correlations of the SZ effects are quite minor,
the impacts on cross-correlations are not negligible.
For example, in the case of the cross-correlation with weak lensing,
even low-mass halos contribute to the signals at large scales \citep{Osato2020}.}
The details of simulations (box size, particle resolution, etc.) and construction of light-cone
outputs are described in \citet{Takahashi2017}.

Since PP requires particle information,
we have run $N$-body simulations using \texttt{Gadget-4} code \citep{Springel2021}
to generate particle snapshots.
First, we compute the transfer function calculated using
Boltzmann code \texttt{CLASS} \citep{Blas2011} and
generated initial condition at the redshift $z = 63$ based on second-order Lagrangian perturbation theory
using \texttt{Gadget-4} code.
The length of the simulation box is $L = 1 \, h^{-1} \, \mathrm{Gpc}$,
the number of particles is $1024^3$, and
the mass of particle is $M_\mathrm{p} = 7.21 \times 10^{10} \, h^{-1} \, \Msun$.
Halos are identified in the on-the-fly manner using \texttt{SubFind} algorithm \citep{Springel2001b},
and only halos whose virial mass is larger than $5 \times 10^{12} \, h^{-1} \, \Msun$ are taken into account
to match with the all-sky halo catalogue.
We simulate 5 independent boxes down to $z = 0$ with distinct initial conditions and
store 10 snapshots for each run.
We then constructed 108 light-cone outputs by stacking 10 snapshots in the line-of-sight direction.
When stacking snapshots, we cut out half of the simulation box, i.e.,
$500 \times 1000 \times 1000 \, (h^{-1} \, \mathrm{Mpc})^3$,
so that our light-cone output covers up to the redshift $z = 3.741$,
which corresponds to the comoving distance of $5000 \, h^{-1} \, \mathrm{Mpc}$.
The contribution from gas beyond the redshift is not incorporated, but such a contribution is negligible.

\begin{figure*}
  \includegraphics[width=\textwidth]{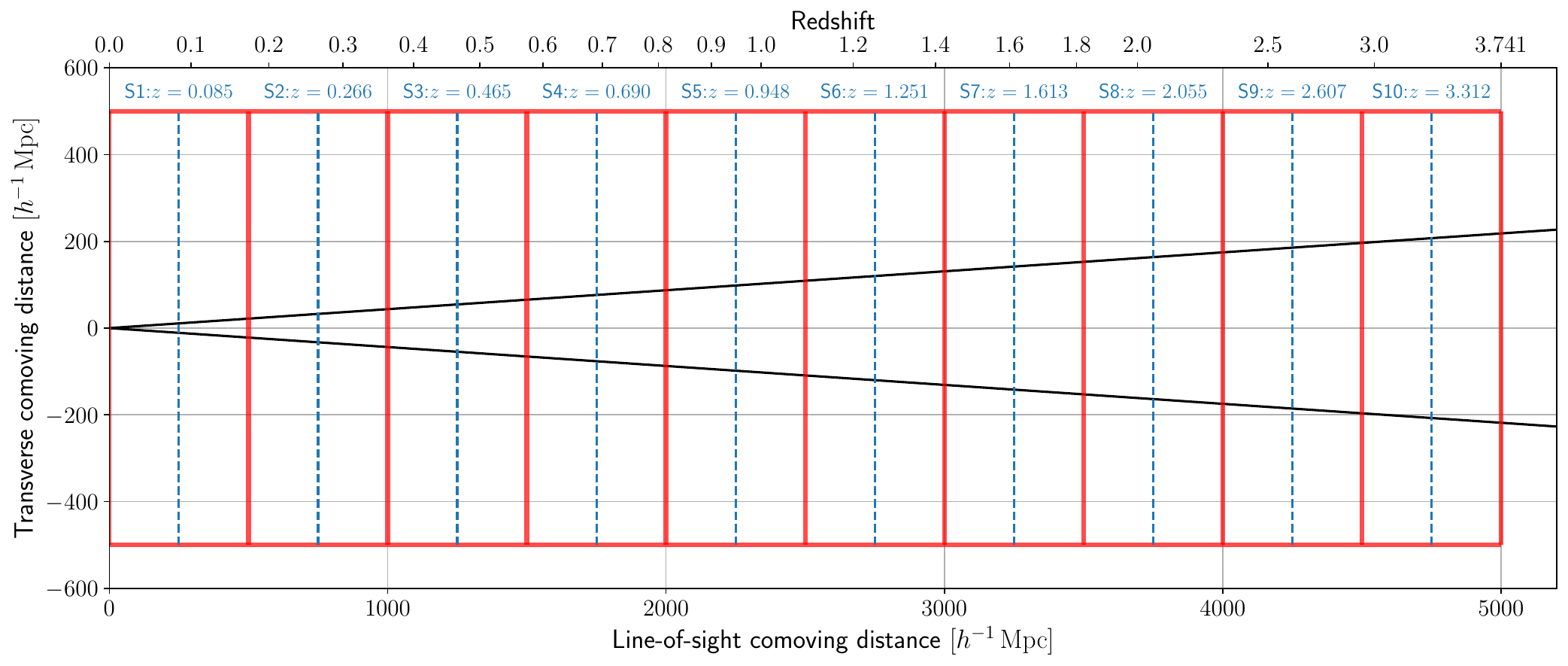}
    \caption{The light-cone output of $N$-body simulation snapshots,
    which is composed of 10 snapshots (labelled as S1, S2, ..., S10).
    The red boxes correspond to
    particle snapshots and blue dashed lines
    show the redshift at which particle snapshots are output.
    The black solid line shows the $5\,\mathrm{deg}$ opening angle.}
    \label{fig:light_cone}
\end{figure*}

The schematic diagram in Figure~\ref{fig:light_cone} illustrates the procedure of our light-cone construction.
To construct pseudo-independent light-cone outputs,
the construction process involves several randomisation processes.
First, for each redshift, we randomly choose one snapshot among 5 realisations and line-of-sight direction
from 6 directions along $(x, y, z, -x, -y, -z)$ axes.
Next, we cut the snapshot by half at the random position and use only one side.
Finally, we randomly rotate and translate all particles
in directions perpendicular to the line-of-sight direction under periodic boundary conditions.
We repeat this procedure for all 10 stacked snapshots and obtain one light-cone output.
In total, we repeat the whole process 108 times
and obtain 108 pseudo-independent mock maps that cover $5 \times 5 \, \mathrm{deg}^2$.

\subsection{Observables of the SZ effects}
In this work, we focus on computing tSZ and kSZ effects and their observables
\citep[see][for reviews]{Birkinshaw1999,Carlstrom2002,Kitayama2014,Mroczkowski2019}.

The tSZ effect is caused by the thermal motion of free electrons.
The inverse Compton scattering between CMB photons and free electrons
are governed by the Kompaneets equation, and then the tSZ temperature variation is given by
\begin{align}
  \frac{\Delta T^\mathrm{tSZ}}{T_\mathrm{CMB}} (\bm{\theta}) &= g_\nu (x) y(\bm{\theta}), \\
  y (\bm{\theta}) &= \frac{\sigma_\mathrm{T}}{m_\mathrm{e} c^2}
  \int_0^{r_\mathrm{rec}} P_\mathrm{e} (l, \bm{\theta}) \mathrm{d} l ,
\end{align}
where $\bm{\theta}$ is the angular coordinate,
$y(\bm{\theta})$ is Compton-$y$ parameter,
$T_\mathrm{CMB} = 2.726 \, \mathrm{K}$ is the CMB temperature,
$\sigma_\mathrm{T}$ is the Thomson scattering cross-section,
$m_\mathrm{e}$ is the electron mass,
$c$ is the speed of light, and $l$ is the physical distance from the observer.
The integration is carried out from the observer ($z = 0$) to
the last-scattering surface $r_\mathrm{rec}$ ($z_\mathrm{rec} \sim 10^3$).
We assume that gas is fully ionised and the electron pressure $P_\mathrm{e}$ is
related to the thermal gas pressure $P_\mathrm{th}$:
\begin{equation}
  P_\mathrm{e} = \frac{\mu}{\mu_\mathrm{e}} P_\mathrm{th} = \frac{2X+2}{5X+3} P_\mathrm{th} ,
\end{equation}
where $\mu = 4/(5X+3)$ is the mean molecular weight,
$\mu_\mathrm{e} = 2/(X+1)$ is the mean molecular weight of electrons, and
$X = 0.76$ is the primordial hydrogen mass fraction.
This variation depends on the observed frequency $\nu$
and the frequency-dependent part $g_\nu (x)$ is given by
\begin{equation}
  g_\nu (x) = x \frac{e^x - 1}{e^x + 1} - 4, \ x = \frac{h \nu}{k_\mathrm{B} T_\mathrm{CMB}} ,
\end{equation}
where $h$ is the Planck constant, and $k_\mathrm{B}$ is the Boltzmann constant.

The kSZ effect is induced by the bulk flow of the electrons.
The kSZ temperature variation is given by
\begin{align}
  \frac{\Delta T^\mathrm{kSZ}}{T_\mathrm{CMB}} (\bm{\theta}) &=
  -b (\bm{\theta}) , \\
  b (\bm{\theta}) &= \frac{\sigma_\mathrm{T}}{c}
  \int_0^{r_\mathrm{rec}}  e^{-\tau (l)} n_\mathrm{e} (l, \bm{\theta}) v_r (l, \bm{\theta}) \mathrm{d}l,
\end{align}
where $n_\mathrm{e}$ is the electron number density and
$v_r$ is the peculiar bulk velocity in the line-of-sight direction
(positive for gas receding from the observer).
We compute the mean optical depth $\tau (z)$ of Thomson scattering
with cosmic baryons up to the distance with the corresponding redshift $z$:
\begin{equation}
  \tau(z) = \sigma_\mathrm{T} \int_0^z \frac{c \mathrm{d}z}{H(z)} (1+z)^2 \bar{n}_\mathrm{e} ,
\end{equation}
where $\bar{n}_\mathrm{e} = \Omega_\mathrm{b} \rho_\mathrm{cr} (z=0) / (\mu_\mathrm{e} m_\mathrm{p})$
is the mean electron density in the present Universe
and $m_\mathrm{p}$ is the proton mass.
The temperature variation due to the kSZ effect has no frequency dependence in contrast to the tSZ effect
and the amplitude of the kSZ signal is smaller than foreground emissions,
which makes the detection by component separation harder.

\subsection{Halo-based pasting}
The first method for computing the SZ observables is the halo-based pasting (HP) algorithm.
According to the halo model, where all matter and gas are associated with halos,
the SZ signals come only from gas embedded within halos,
and the diffuse gas's contribution outside the halos is assumed to be zero.

\begin{figure}
  \includegraphics[width=\columnwidth]{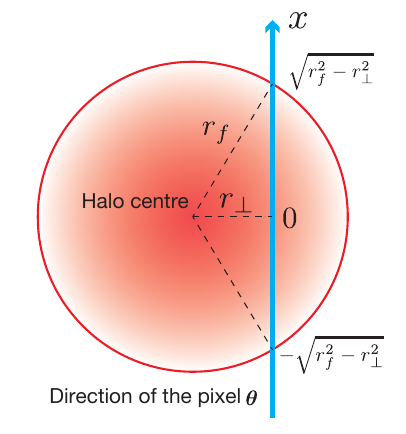}
    \caption{A schematic picture for HP map making in the case of a single halo.
    The blue line corresponds
    to the direction of the pixel $\bm{\theta}$.}
    \label{fig:schematic_HP}
\end{figure}

First, for each halo (labelled as $h$) in a halo catalogue,
we compute the electron pressure profile $P_{\mathrm{e}, h} (r)$
and the electron number density profile $n_{\mathrm{e}, h} (r)$
by solving the ICM model described in Section~\ref{sec:model}
with the halo mass, the redshift, and the concentration parameter provided in the halo catalogue.
Then, for every pixel of a map at the angular position $\bm{\theta}$ around each halo center,
we integrate the SZ observables in the line-of-sight direction:
\begin{align}
  y^\mathrm{HP} (\bm{\theta}) &= \sum_{h:\text{halo}} \frac{\sigma_\mathrm{T}}{m_\mathrm{e} c^2}
  2 \int_{0}^{\sqrt{r_f^2 - r_\perp^2}}
  P_{\mathrm{e},h} \left( \sqrt{r_\perp^2+x^2} \right) \mathrm{d}x , \\
  b^\mathrm{HP} (\bm{\theta}) &= \sum_{h:\text{halo}} \sigma_\mathrm{T} \frac{v_h}{c} e^{-\tau_h}
  2 \int_{0}^{\sqrt{r_f^2 - r_\perp^2}}
  n_{\mathrm{e},h} \left( \sqrt{r_\perp^2+x^2} \right) \mathrm{d}x ,
\end{align}
where $v_h$ is the line-of-sight bulk velocity of the halo
(i.e., the mean velocity of particles within the virial radius),
$\tau_h$ is the optical depth from the observer to the halo centre,
and $x$ is the coordinate where the coordinate axis is parallel
to the direction of the interested pixel
and the origin is the perpendicular foot from the halo centre to the axis.
The summation runs over halos whose impact parameter $r_\perp$
is smaller than $r_f$.
Figure~\ref{fig:schematic_HP} shows the schematic diagram
in the case of a single halo where the line-of-sight integration through the halo is performed at the pixel location $\bm{\theta}$.

\subsection{Particle-based pasting}
The second method for computing the SZ observables is the particle-based pasting (PP) algorithm,
where all gas (inside and outside of halos) can contribute to the SZ signals.

To do this, we follow the map-making procedure by \citet{Roncarelli2007,Ursino2010},
which utilised the hydrodynamical simulations to produce mock maps.
First, we consider contributions from particles that belong to each halo.
Similarly to HP, we solve the ICM model to obtain the pressure and number density profiles.
The contribution from each particle (labelled as $i$) is given by
\begin{align}
  y_i^\mathrm{PP,h} &= \frac{\sigma_\mathrm{T}}{m_\mathrm{e} c^2} P_{\mathrm{e},i} V_i L_\mathrm{pix}^{-2} , \\
  b_i^\mathrm{PP,h} &= \sigma_\mathrm{T} \frac{v_i}{c} e^{-\tau_i} n_{\mathrm{e}, i} V_i L_\mathrm{pix}^{-2},
\end{align}
where $L_\mathrm{pix}$ is the physical pixel size at the particle position.
The volume of each particle is given by $V_i = M_\mathrm{p} / \rho_\mathrm{tot} (r_i)$,
where $r_i$ is the distance from the halo centre and
$\rho_\mathrm{tot} (r_i)$ is the density computed from the NFW density profile (Eq.~\ref{eq:NFW_profile}).

\begin{figure}
  \includegraphics[width=\columnwidth]{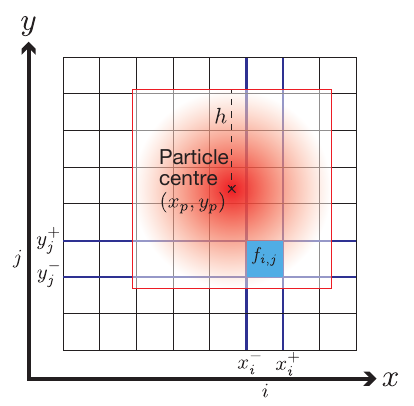}
    \caption{A schematic picture for PP map making in the case of a single particle.
    The red square is the particle-centred square with a length on a side
    twice the smoothing length $2h$
    and pixels overlapping this square have a non-zero weight.
    The blue pixel corresponds to the target for the assignment, and the weight of this pixel is $f_{i,j}$.}
    \label{fig:schematic_PP}
\end{figure}

We assume that each particle has a finite spatial extent characterised
by the smoothed particle hydrodynamics (SPH) kernel \citep{Gingold1977}:
\begin{equation}
  W_\mathrm{SPH} (x) = \frac{8}{\pi h^3}
  \begin{cases}
    1 - 6 x^2 +6 x^3 , & (0 \leq x < 1/2) \\
    2 (1-x)^3 , & (1/2 \leq x < 1) \\
    0 , & (1 \leq x)
  \end{cases}
  \label{eq:SPH_kernel}
\end{equation}
where $x = r/h_i$ and $h_i$ is the smoothing length.
The smoothing length is determined so that the effective number of particles within the smoothing length
should be close to
$N_\mathrm{ngb} = 64$ \citep[see][for more details]{Springel2021}.\footnote{Although it is possible
to use the SPH density to estimate the particle volume $V_i$,
the SPH density overestimates the true density \citep{Pelupessy2003}.
The SZ signal is proportional to the particle volume,
and thus, a higher density may lead to a lower signal at the outskirts of halos.
Thus, we opt for using the SPH kernel only to represent the spatial extent.}
Whereas the SPH kernel represents the 3D spatial extent,
the kernel for the 2D spatial extent is required for map making.
Strictly speaking, the 2D kernel can be obtained by projecting the 3D kernel
but the projected 2D kernel contains complex expressions and is time-consuming to compute.
Thus, instead of the projected 2D kernel, we keep using the expressions of the 3D version
(Eq.~\ref{eq:SPH_kernel}) by plugging the projected 2D radius instead of the 3D radius.
Since the shape of both the 2D projected kernel and the original 3D kernel is similar to
the normal distribution, this approximation does not have a significant impact on the final results.
Furthermore, the kernel can be approximated as the product
of the kernels along 2D axes, i.e., $W_\mathrm{SPH} (r) \simeq W_\mathrm{SPH}(x) W_\mathrm{SPH}(y)$ \citep{Ursino2010}
and this approximation only affects the small-scale feature, which is well below the resolution of our simulations.
This factorization enables the integration along the axis in advance,
and then, for each pixel, we assign the signal with the weight $f_{i, j}$:
\begin{align}
  f_{i, j} &= \big\{ w_\mathrm{SPH} \big[ (x_i^+ - x_p)/h \big] - w_\mathrm{SPH} \big[ (x_i^- - x_p)/h \big] \big\}
  \nonumber \\
  & \times \big\{ w_\mathrm{SPH} \big[ (y_j^+ - y_p)/h \big] - w_\mathrm{SPH} \big[ (y_j^- - y_p)/h \big] \big\} ,
\end{align}
where $(x_p, y_p)$ is the physical coordinates of the particle centre,
and $x_i^-$ and $y_j^-$ ($x_i^+$ and $y_j^+$) are lower (upper) edges of the pixel $(i, j)$,
and the projected kernel function $w_\mathrm{SPH}$ is given as
\begin{align}
  w_\mathrm{SPH} (x) &\equiv \int_{-\infty}^x W_\mathrm{SPH} (|x'|) \mathrm{d} x'
  \bigg/ \int_{-\infty}^\infty W_\mathrm{SPH} (|x'|) \mathrm{d} x'
  \nonumber \\
  & =
  \begin{cases}
    0 , & (x < -1) \\
    \frac{2}{3} (1+x)^4 , & (-1 < x \leq -1/2) \\
    \frac{4}{3} \left( \frac{3}{8} + x - 2 x^3 - \frac{3}{2} x^4 \right) , & (-1/2 < x \leq 0) \\
    \frac{4}{3} \left( \frac{3}{8} + x - 2 x^3 + \frac{3}{2} x^4 \right) , & (0 < x \leq 1/2) \\
    -\frac{2}{3} (1-x)^4 + 1 , & (1/2 < x \leq 1) \\
    1 . & (x > 1)
  \end{cases}
  \label{eq:integrated_kernel}
\end{align}
Since the weight is automatically normalized, i.e.,
\begin{equation}
\sum_{i, j} f_{i, j} = 1 ,
\end{equation}
there is no need for normalisation, which speeds up the assignment of signals.
Figure~\ref{fig:schematic_PP} illustrates the assignment of the signal from a single particle.
Note that it is possible to assign the contributions from particles
with the nearest grid point or cloud-in-cell algorithms.
These methods yield results consistent with the results assigned to the SPH kernel
at large scales \citep{McCarthy2014}
but cannot capture the damping behaviour of
angular power spectra at small scales ($\ell \gtrsim 10^4$).

For field particles, the contribution from a particle is given by
\begin{align}
  y_i^\mathrm{PP,f} &= 0 ,
  \label{eq:field_y} \\
  b_i^\mathrm{PP,f} &= \sigma_\mathrm{T} \frac{v_i}{c} \frac{M_\mathrm{p}}{\mu_\mathrm{e} m_\mathrm{p}}
  e^{-\tau_i} L_\mathrm{pix}^{-2}.
  \label{eq:field_b}
\end{align}
Then, the signal is assigned to pixels according to the projected SPH kernel
in the same way as halo particles.
The contribution from field particles to the tSZ signal is assumed to be zero
but the tSZ effect from gas in filamentary structures has been reported \citep{deGraaff2019}.
Once the pressure profile in filaments is given,
it is straightforward to take into account the contributions
by modifying the field term $y_i^\mathrm{PP,f}$ (Eq.~\ref{eq:field_y}).

\section{Theory}
\label{sec:theory}
In this section, we review theoretical models to predict angular power spectra of tSZ and kSZ effects.

\subsection{tSZ power spectrum}
For the Compton-$y$ power spectrum, the dominant source of the signal
is hot electrons in dark matter halos, and the contribution of the diffuse component of the cosmic web
lying outside of dark matter halos is relatively minor.
Thus, the halo model prescription is justified.
The power spectrum based on the halo model \citep{Komatsu1999,Komatsu2002} is
given by the sum of the 1-halo $C^\mathrm{1h}_y$ and 2-halo $C^\mathrm{2h}_y$ terms:
\begin{align}
  C_y (\ell) &= C^\mathrm{1h}_y (\ell) + C^\mathrm{2h}_y (\ell), \\
  C^\mathrm{1h}_y (\ell) &= \int_0^{z_\mathrm{max}} \mathrm{d}z \frac{\mathrm{d}^2 V}{\mathrm{d}z \mathrm{d}\Omega}
  \int \mathrm{d}M \frac{\mathrm{d}n_\mathrm{h}}{\mathrm{d}M} (M, z) |\tilde{y} (\ell; M, z)|^2 , \\
  C^\mathrm{2h}_y (\ell) &= \int_0^{z_\mathrm{max}} \mathrm{d}z \frac{\mathrm{d}^2 V}{\mathrm{d}z \mathrm{d}\Omega}
  P_\mathrm{L} \left( k = \frac{\ell + 1/2}{D_A (z)}, z \right) \nonumber \\
  & \times \left[ \int \mathrm{d}M \frac{\mathrm{d}n_\mathrm{h}}{\mathrm{d}M} (M, z)
  b_\mathrm{h} (M, z) \tilde{y} (\ell; M, z) \right]^2 ,
\end{align}
where $\mathrm{d}^2 V/\mathrm{d}z \mathrm{d}\Omega = \chi^2 c/H(z)$
is the comoving volume per redshift and solid angle,
$\chi (z)$ is the comoving distance,
$\mathrm{d}n_\mathrm{h} / \mathrm{d}M (M, z)$ is the halo mass function,
$b_\mathrm{h} (M, z)$ is the halo bias,
$P_\mathrm{L} (k, z)$ is the linear matter power spectrum, and
$D_A (z)$ is the angular diameter distance.
The maximum redshift is $z_\mathrm{max} = 3.741$, which corresponds
to the edge of the light-cone output.\footnote{This value is slightly different
from the maximum redshift of the all-sky catalogue $z = 3.65$, but its impact on power spectra is negligible.}
For the halo mass function and the halo bias,
we adopt fitting formulas calibrated with $N$-body simulations
in \citet{Tinker2008} and \citet{Tinker2010}, respectively.
For the mass integration, virial mass $M_\mathrm{vir}$ is used as a halo mass
and the integration range is $[5 \times 10^{12}, 10^{16}] \, h^{-1} \, \Msun$.
The halo mass function and the halo bias are computed using the fitting formulas,
after converting $M_\mathrm{vir}$ to $M_{200\mathrm{b}}$
through NFW profile.\footnote{$M_{200\mathrm{b}}$ is the halo mass defined
with respect to the enclosed mass with the mean density of $200$ times the mean matter density of the Universe.}
The Fourier transform of 3D Compton-$y$ profile $\tilde{y} (\ell; M, z)$ is given by
\begin{equation}
  \tilde{y} (\ell; M, z) = \frac{4 \pi R_s}{\ell_s^2}
  \frac{\sigma_\mathrm{T}}{m_\mathrm{e} c^2}
  \int_0^\infty \!\! \mathrm{d}x \, x^2 P_\mathrm{e} (x)
  \frac{\sin [(\ell+1/2)x/\ell_s]}{(\ell+1/2)x/\ell_s} ,
\end{equation}
where $P_\mathrm{e}$ is the free electron pressure profile computed with the ICM model
(Section~\ref{sec:model}), $x = r/R_s$, $\ell_s = D_A (z)/R_s$, and $R_s$ is the arbitrary scale radius.

\subsection{kSZ power spectrum}
Unlike the tSZ effect, the kSZ signal involves significant contributions
from diffuse gas outside of dark matter halos.
Therefore, extra care is needed when applying the halo model prescription
for modelling the kSZ effect.
The kSZ effect vanishes at the leading order of density perturbation
due to the cancellation of opposite velocities,
but the second-order perturbation leads to a non-zero power spectrum at sub-arcminute scales.
This effect was first discussed in \citet{Ostriker1986,Vishniac1987}
and now referred to as Ostriker--Vishniac (OV) power spectrum.
Subsequent studies \citep{Ma2002,Shaw2012,Park2016} address
the extension of the OV power spectrum by incorporating nonlinear treatments.

At small scales, the kSZ power spectrum is sourced by the transverse mode given by
\begin{align}
  C_b (\ell) =
  \frac{1}{2} \left( \frac{\sigma_\mathrm{T} \bar{n}_\mathrm{e}}{c} \right)^2
  & \int_0^{z_\mathrm{max}} \frac{c \mathrm{d}z}{H(z)} \frac{(1+z)^4}{\chi^2 (z)}
  \nonumber \\
  & \times P_{q_\perp} \left( k = \frac{\ell+1/2}{D_A(z)}, z \right)
  \exp [-2\tau (z)]  ,
\end{align}
where the transverse momentum power spectrum $P_{q_\perp} (k, z)$ is given by
\begin{align}
  P_{q_\perp} (k, z) =
  (a H f)^2 &\int \frac{\mathrm{d}^3 k'}{(2 \pi)^3} \nonumber \\
  & \times \left[ \frac{1-\mu^2}{k'^2} P_{\de \de} (|\bm{k}-\bm{k}'|, z)
  P_{\theta \theta} (k', z) \right. \nonumber \\
  & \left. - \frac{1-\mu^2}{|\bm{k}-\bm{k}'| k'} P_{\de \theta} (|\bm{k}-\bm{k}'|, z)
  P_{\de \theta} (k', z) \right] .
  \label{eq:perp_power}
\end{align}
This expression does not depend on the direction of $\bm{k}$, and
we take the direction along $z$-direction, i.e., $\bm{k} = (0, 0, k)$.
$\mu$ is the directional cosine between
wavevectors $\bm{k}$ and $\bm{k}'$,
$\theta \equiv -\nabla \cdot \bm{v} / (aHf)$ is the normalised velocity divergence,
$a = 1/ (1+z)$ is the scale factor,
$f = \mathrm{d} \ln D_+ (a) / \mathrm{d} \ln a$ is the linear growth rate,
$D_+ (a)$ is the linear growth factor normalized as $D_+ (a = 0) = 1$,
$P_{\de \de} (k, z)$ and $P_{\theta \theta} (k, z)$ are the auto-power spectra of
electron density and normalised velocity divergence, respectively,
and $P_{\de \theta} (k, z)$ is the cross-power spectrum of
electron density and normalised velocity divergence.
When the electron density field follows
the matter density field ($\delta_\mathrm{e} = \delta$)
in the linear regime ($\delta = \theta$),
the expression is given by
\begin{equation}
  P_{q_\perp}^\mathrm{OV} (k, z) = (aHf)^2 \int \frac{\mathrm{d}^3 k'}{(2 \pi)^3}
  P_\mathrm{L} (|\bm{k}-\bm{k}'|, z) P_\mathrm{L} (k', z) I (k, k'),
\end{equation}
where
\begin{equation}
  I (k, k') = \frac{k (k - 2k' \mu) (1-\mu^2)}{k'^2 (k^2 + k'^2 - 2 k k' \mu)} .
\end{equation}
The angular power spectrum at the linear order is referred to as the OV power spectrum.

Next, we consider the nonlinear extension of the OV power spectrum.
We make use of the fitting formula of electron density power spectrum $b_\mathrm{e} (k, z)$
calibrated using hydrodynamical simulations \citep[Eq.~21 of][]{Takahashi2021}
and the fitting formulas for velocity divergence power spectra,
$b_{\delta \theta} (k)$ and $b_{\theta \theta} (k)$
\citep[Eqs.~24 and 25 of][]{Hahn2015}\footnote{Note that we normalise the velocity divergence field
while \citet{Hahn2015} uses the unnormalised velocity divergence field.
The fitting formulas of \citet{Hahn2015} ($b_{\theta}^{(1)}$ and $b_{\theta}^{(2)}$)
are related to $b_{\delta \theta}$ (Eq.~\ref{eq:fitting_dt}) and $b_{\theta \theta}$ (Eq.~\ref{eq:fitting_tt}) as
$b_{\delta \theta} = b_{\theta}^{(1)}$ and $b_{\theta \theta} = \left( b_{\theta}^{(2)} \right)^2$.}:
\begin{align}
  P_{\de \de} (k, z) &= b_\mathrm{e}^2 (k, z) P_\mathrm{NL} (k, z), \\
  \label{eq:fitting_dt}
  P_{\de \theta} (k, z) &= b_\mathrm{e} (k, z)
  b_{\delta \theta} (k) P_\mathrm{NL} (k, z), \\
  \label{eq:fitting_tt}
  P_{\theta \theta} (k, z) &= b_{\theta \theta} (k)P_\mathrm{NL} (k, z),
\end{align}
where $P_\mathrm{NL} (k, z)$ is the nonlinear matter power spectrum
computed using the \textit{halofit} prescription \citep{Smith2003}
with updated parameters of \citet{Takahashi2012}.
By substituting these nonlinear power spectra in the momentum power spectrum (Eq.~\ref{eq:perp_power}),
we compute the nonlinear kSZ power spectrum.

The longitudinal mode is subdominant at small scales, while it dominates over
the transverse mode on the large scales ($\ell \lesssim 100$) \citep{Park2016,Alvarez2016}.
Since this mode is only important at scales larger than the one
where observations are available ($100  \lesssim \ell \lesssim 10^4$) and
the finite size of the simulation box may lead to a lack of large-scale modes,
we do not include the longitudinal mode contribution in this work.

\section{Results}
\label{sec:results}
In this section, we present mock HP and PP maps and briefly discuss their properties.

\subsection{Flat-sky HP and PP maps}
\label{sec:flat-HP-PPmaps}
\begin{figure*}
  \includegraphics[width=\columnwidth]{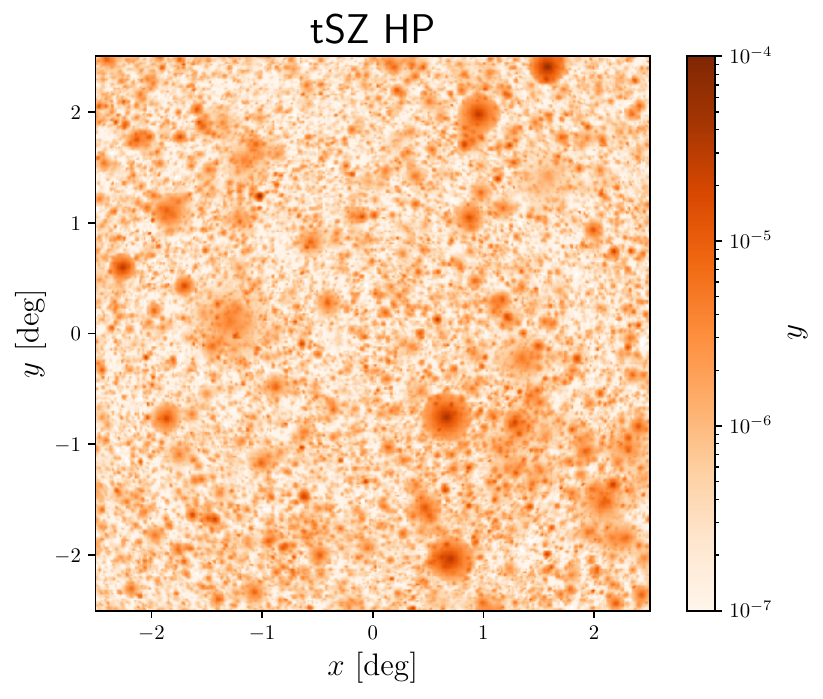}
  \includegraphics[width=\columnwidth]{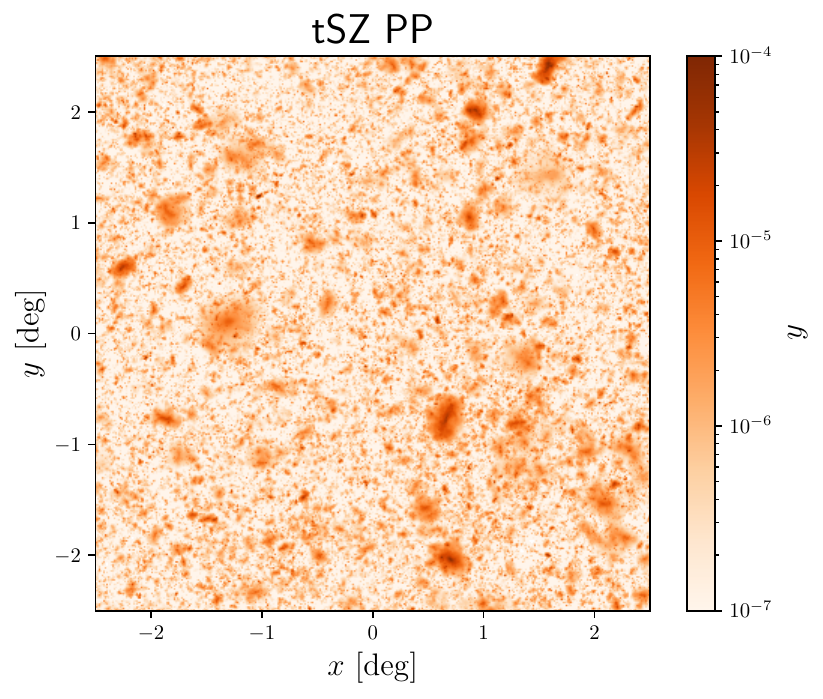}
  \includegraphics[width=\columnwidth]{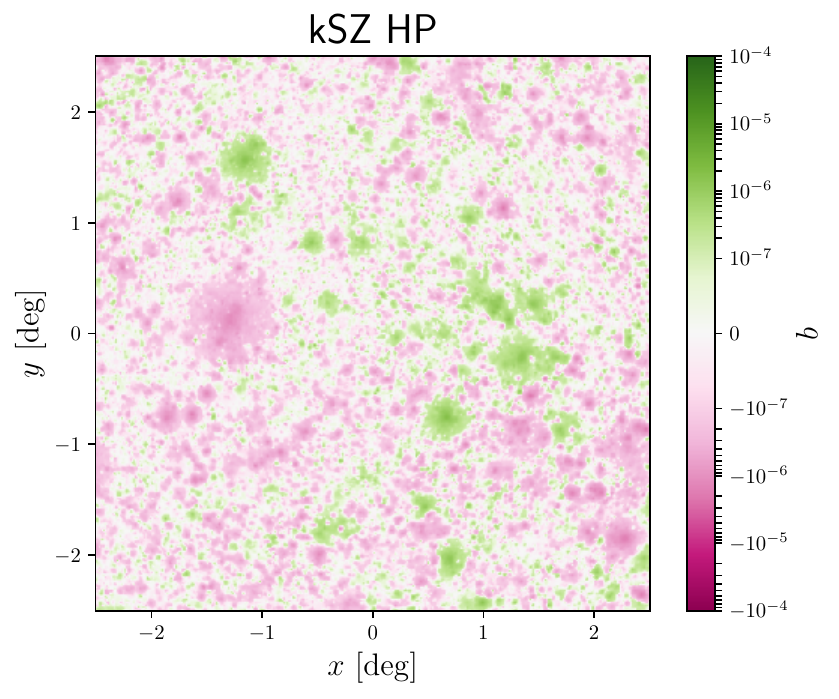}
  \includegraphics[width=\columnwidth]{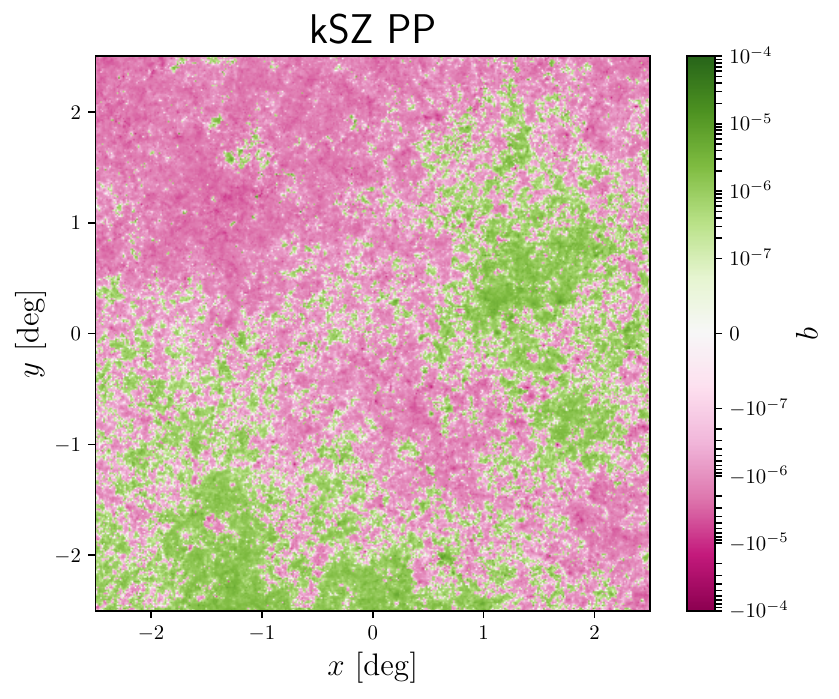}
    \caption{
    A $5 \times 5 \, \mathrm{deg}^2$ mock tSZ map (top) and kSZ map (bottom)
    from the same simulation outputs based on the HP (left) and PP (right) pasting methods.
    }
    \label{fig:HP_PP_ybmap}
\end{figure*}

Figure~\ref{fig:HP_PP_ybmap} shows the comparison of HP (left panels) and PP (right panels)
methods for computing tSZ (top panels) and kSZ maps (bottom panels) among 108 realisations.
The maps are pixellated with regular grids, and the number of grids on a side is $4096$,
where the pixel size is $5 \, \mathrm{deg} / 4096 = 4.39 \, \mathrm{arcsec}$.
The PP tSZ map looks quite similar to the HP tSZ map.
However, the signal is not perfectly circular symmetric because
the gas is painted onto particles whose distribution is not spherically symmetric.
Though the ICM model assumes spherical symmetry,
we can incorporate the asphericity of halos in PP maps to some extent.\footnote{The asphericity of halos
can be incorporated by adopting triaxial gas density and pressure profiles within the HP framework (Lau et al., in prep.).}
The spherical symmetry adopted in HP smears out small-scale structures
and makes the overall signals more diffuse than the corresponding PP map.

\begin{figure*}
  \includegraphics[width=\textwidth]{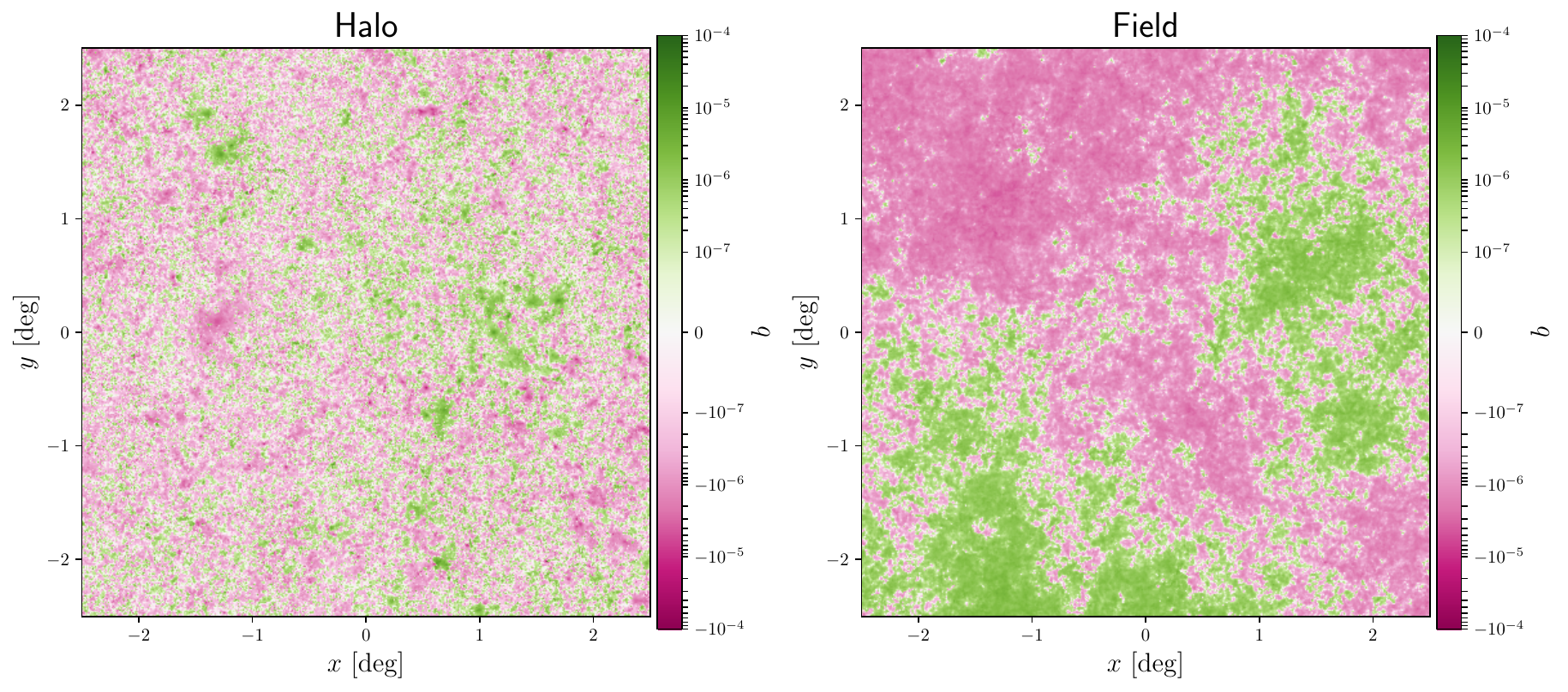}
    \caption{A $5 \times 5 \, \mathrm{deg}^2$ mock PP kSZ map showing the contributions
    from halo (left panel) and field components (right panel), respectively}
    \label{fig:PP_halo_field_bmap}
\end{figure*}

One of the most notable distinctions lies in the kSZ maps created using HP and PP
shown in the bottom left and bottom right panels of Figure~\ref{fig:HP_PP_ybmap},
because of the significant contribution of the kSZ signal from the gas outside
of the virialization region in dark matter halos. To quantify the relative
contributions from halos and fields, we show the kSZ maps created using the PP method
by including only halo or field particles in the left and right panels
of Figure~\ref{fig:PP_halo_field_bmap}, respectively.
Since the HP kSZ map does not include field contribution, it is fair to compare
it with the halo particle-only PP map to the left panel of Figure~\ref{fig:PP_halo_field_bmap}.
For kSZ maps, the critical difference between the HP and PP methods is that the HP method
uses the averaged halo velocity. In contrast, the PP method uses full particle velocity
information, which enables the PP method to capture some of the small-scale kinematics
better than the HP method. Therefore, the small-scale feature is more manifest in the PP map,
while such a feature is smeared out in the HP map.

\subsection{All-sky HP maps}
\begin{figure*}
  \includegraphics[width=\textwidth]{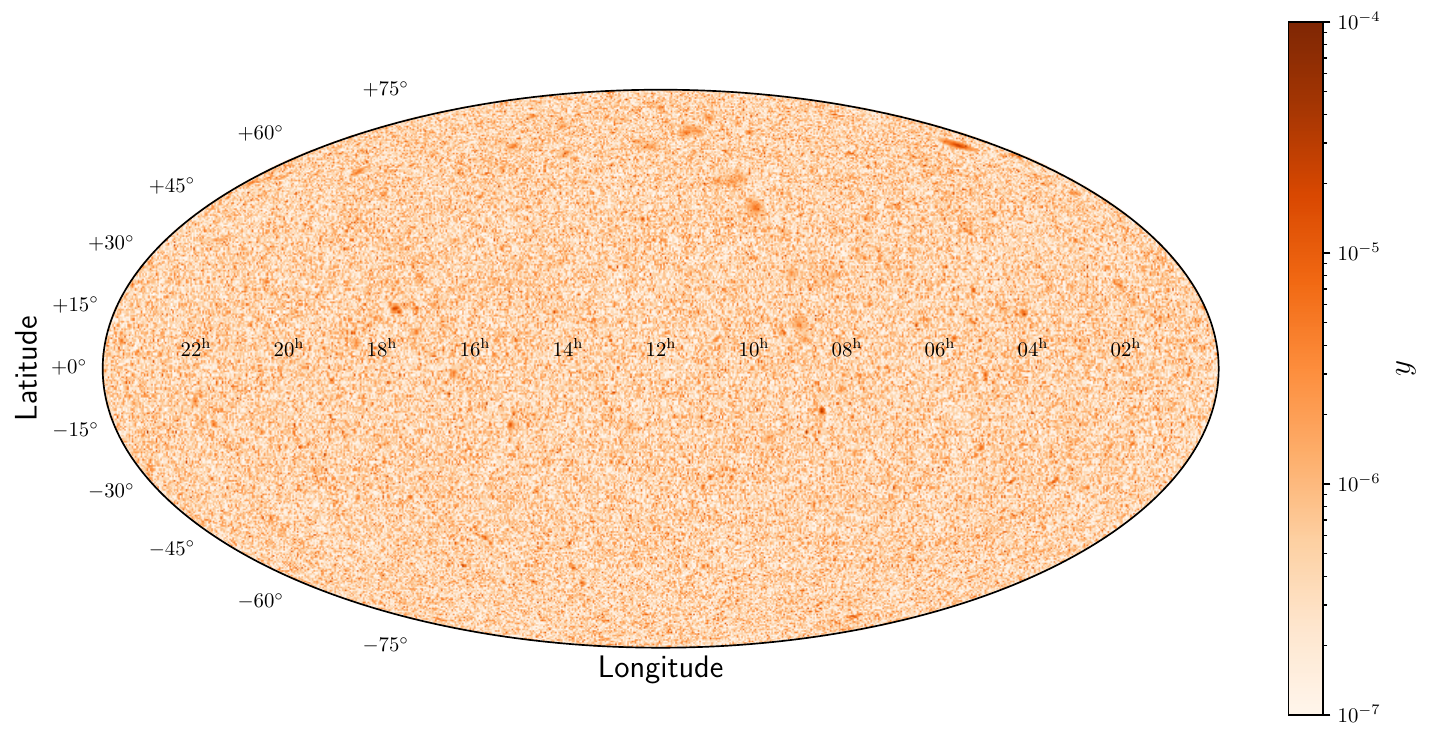}
  \includegraphics[width=\textwidth]{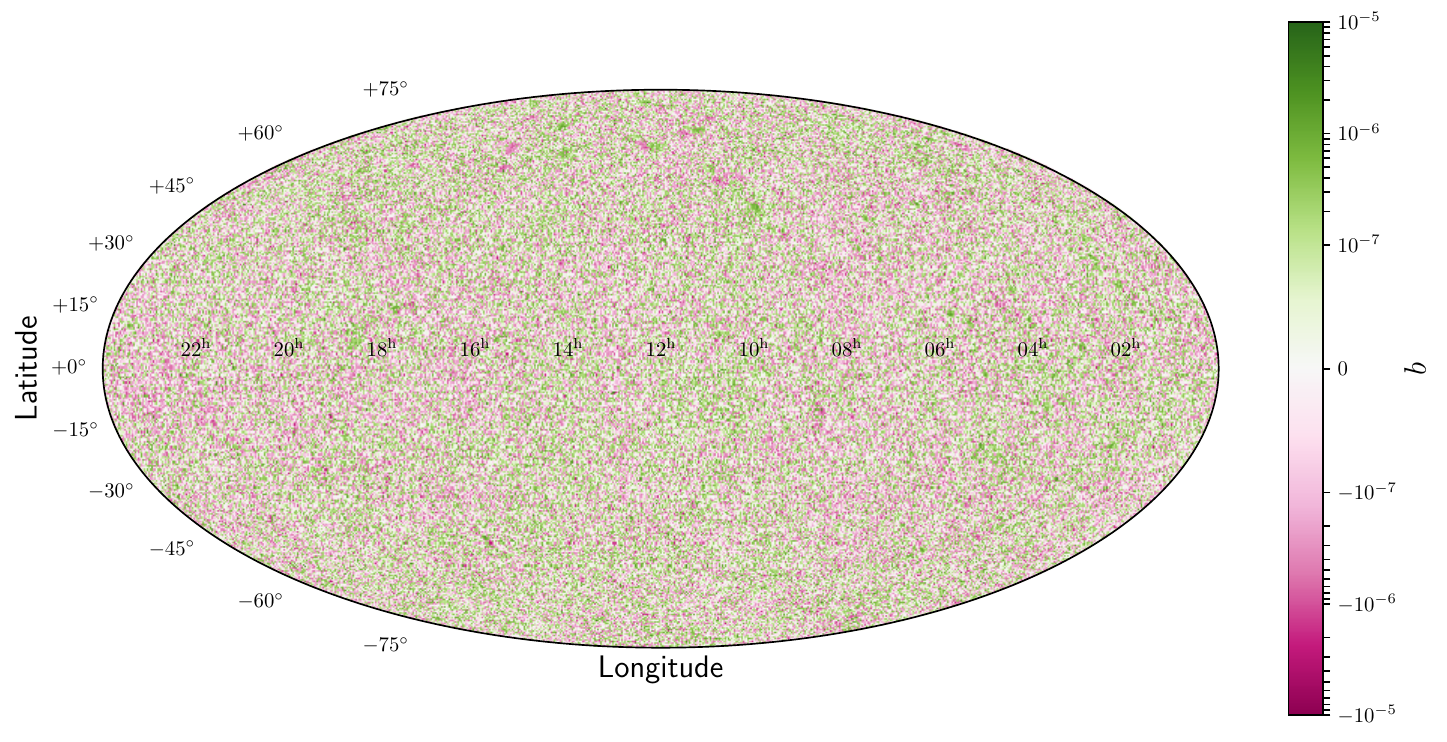}
    \caption{An all-sky mock HP tSZ map (top) and kSZ map (bottom).}
    \label{fig:HP_ybmap}
\end{figure*}

Since the HP algorithm is computationally more efficient than the PP algorithm and requires only a halo catalogue,
we generate 108 all-sky mock tSZ and kSZ maps by applying the HP method on the halo catalogue created
from the simulations described in Section~\ref{sec:simulation}.
First, we show an example of all-sky HP mock tSZ and kSZ maps among 108 realisations
in the top and bottom panels of Figures~\ref{fig:HP_ybmap}, respectively.
We employ the \texttt{Healpix} pixellation \citep{Gorski2005}
with $N_\mathrm{side} = 8192$, which corresponds to the pixel area of $0.184 \, \mathrm{arcmin}^2$
and the effective pixel length of $25.8 \, \mathrm{arcsec}$.
Peaks, which correspond to massive halos, can be visually identified in the tSZ and kSZ maps.

\begin{figure*}
  \includegraphics[width=\columnwidth]{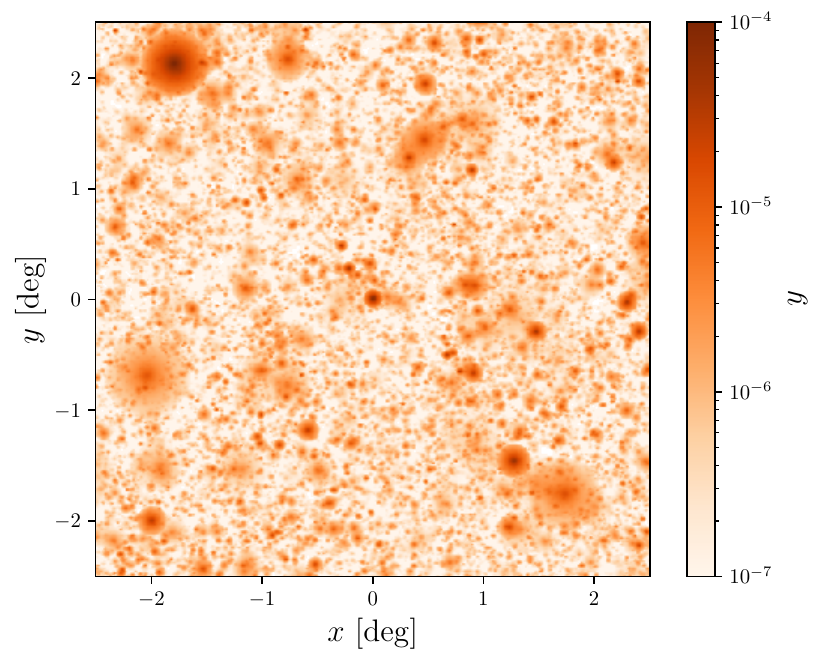}
  \includegraphics[width=\columnwidth]{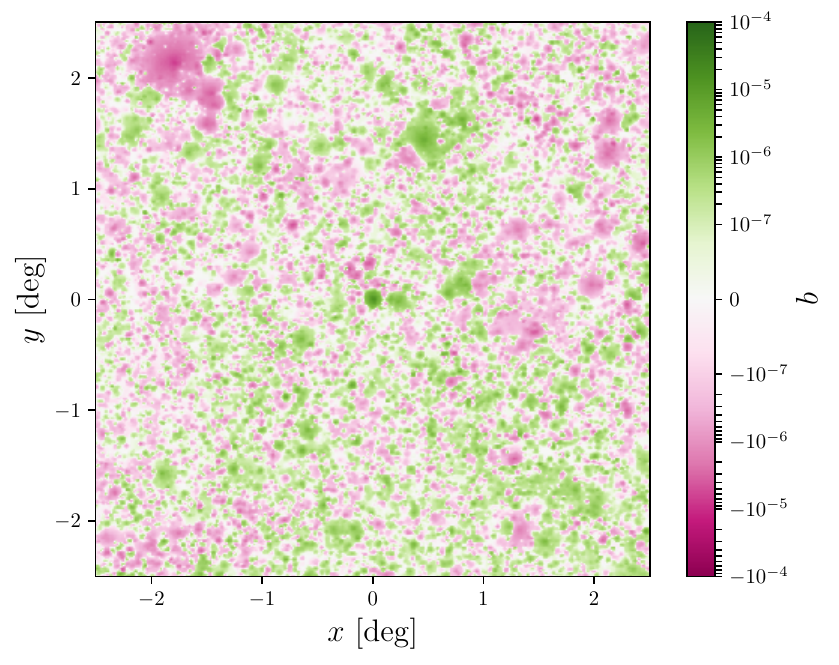}
    \caption{A zoom up ($5 \times 5 \, \mathrm{deg}^2$) view of the HP tSZ map (left) and kSZ map (right)
    centred at one of Compton-$y$ peaks.}
    \label{fig:HP_ybmap_zoom}
\end{figure*}

Next, we show an example of a smaller ($5 \times 5 \, \mathrm{deg}^2$) patch of these maps
to zoom into the region centered at one of the Compton-$y$ peaks (the peak value is $y = 1.1\times 10^{-4}$)
in Figure~\ref{fig:HP_ybmap_zoom}.
Since spherical symmetry is assumed in the HP model,
spherical clouds can be seen in both tSZ and kSZ maps.
Since the HP algorithm only considers contributions from halos,
there are strong correlations between tSZ and kSZ maps.
As we have seen in the comparison of HP and PP maps in Section~\ref{sec:flat-HP-PPmaps},
the kSZ signal of an HP map looks smoother than that of a PP map,
because the bulk velocity is the mean velocity within a halo, and the small-scale kinematics is smeared out.

\subsection{Power spectra}
The power spectra are the fundamental statistics in SZ measurements
for constraining cosmology and astrophysics of ICM,
and estimation of the covariance matrix of power spectra is one of the primary applications for mock SZ maps.
We, therefore, measure the power spectra of mock all-sky HP and flat-sky PP maps
and compare them to theoretical predictions.
We have checked that the power spectra of flat-sky HP and PP maps are almost identical. Thus, we show only the results of flat-sky PP maps.
Due to the difference in the sky coverage,
we measure power spectra for the multipole ranges of $2 < \ell < 10^4$ for all-sky HP maps
and $10^2 < \ell < 5 \times 10^4$ for flat-sky PP maps.
For all-sky maps, we measure the power spectrum with \texttt{anafast} routine in \texttt{healpy} module
and deconvolve the window function of \texttt{Healpix} pixellation.
For flat-sky maps, we measure the power spectrum with a fast Fourier transform, and
the multipole bin is log-equally sampled in the range of $10^2 < \ell < 10^5$
with $30$ bins.

\begin{figure}
  \includegraphics[width=\columnwidth]{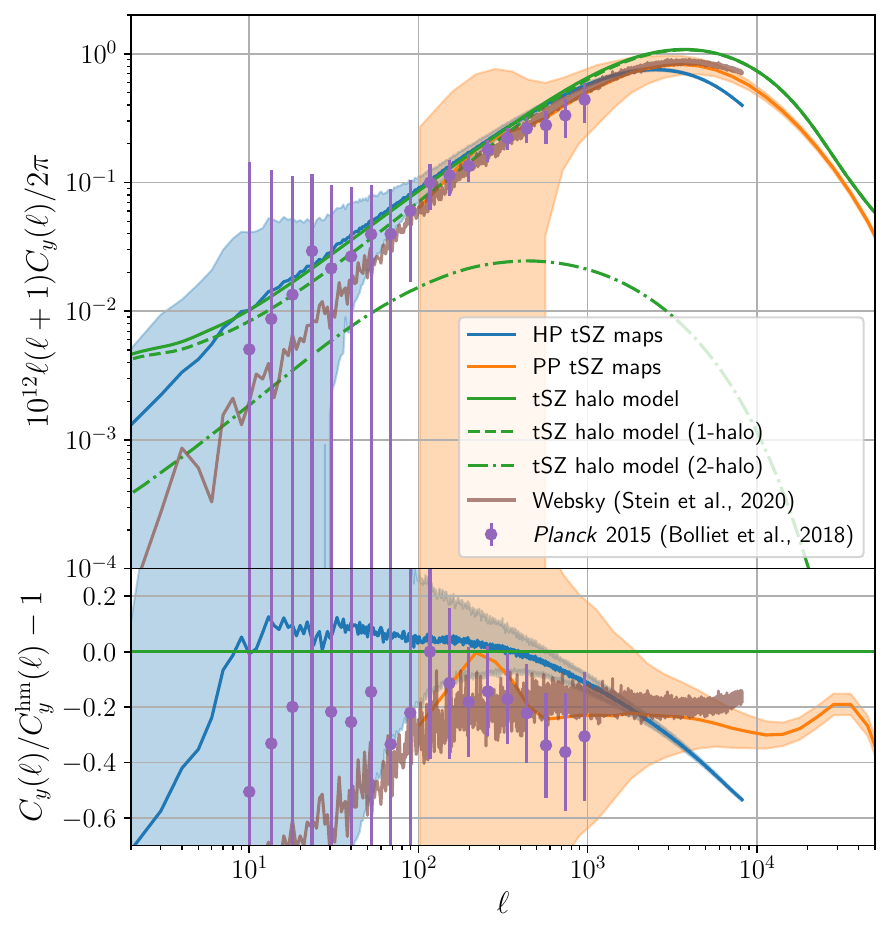}
    \caption{The angular power spectra of mock all-sky HP and flat-sky PP tSZ maps.
    The solid lines and shaded regions correspond to
    the mean and the standard deviation of 108 HP and PP maps, respectively.
    Note that the magnitude of the errors is different because the sky coverage is different
    (all-sky for HP maps and $5 \times 5\,\mathrm{deg}^2$ for PP maps)
    and power spectra of PP maps are binned spectra.
    For comparison, the halo model prediction, Websky mock all-sky simulation \citep{Stein2020},
    and the re-analysis result of \textit{Planck} 2015 measurement \citep{Bolliet2018} are also shown.}
    \label{fig:cl_tSZ}
\end{figure}

Figure~\ref{fig:cl_tSZ} shows power spectra of tSZ HP and PP maps
along with the halo model prediction by the Websky mock all-sky map \citep{Stein2020}
and the result of the re-analysis of \textit{Planck} 2015 measurement \citep{Bolliet2018} for comparison.
At the intermediate multipole range $10 < \ell < 3000$,
where both HP and PP results are less susceptible to resolution effects,
the power spectra of PP maps are significantly lower than those of HP maps.
In terms of comparison with the halo model, HP maps are consistent with the prediction up to the resolution limit
$\ell \simeq 1000$ at the level of $10\%$ since the halo model assumption holds for HP maps.
The discrepancy on large scales can be attributed to the halo mass function and halo concentration parameter;
these two components are directly measured in simulations and are more accurate in HP maps.
On the other hand, the spectra of PP maps are substantially lower than HP maps and halo model prediction by $20\%$,
because PP maps incorporate halos' asphericity, which is missing in the halo model assumption.
It has been known that the shape of halos has an impact on scaling relations of integrated Compton-$y$-mass \citep{Battaglia2012a}
and temperature-mass \citep{Chen2019}.
In comparison with \textit{Planck} 2015 measurement \citep{Bolliet2018},
PP maps are more consistent with the halo model prediction and HP maps.
It has been recognized that the halo model prediction with best-fit cosmological parameters from CMB temperature and polarisation results yields a larger power spectrum compared with the observations \citep{Salvati2018},
which is related to the so-called $\sigma_8$ tension.
The inclusion of the asphericity of halos, as in our PP method,
mitigates the tension by lowering the amplitude of the tSZ power spectrum.
We also show the comparison with the Websky mock tSZ map \citep{Stein2020}.
Since the Websky simulation adopts different cosmological parameters, we scale the power spectrum with the relation $C_y \propto \sigma_8^{8.1} \Omega_\mathrm{m}^{3.2} h^{-1.7}$ \citep{Bolliet2018}
to match the cosmological parameters used to generate our HP and PP maps.
While the map-making algorithm in the Websky simulations is similar to our HP method,
the resultant power spectrum is significantly lower than the halo model prediction and results of HP maps.
That is because the Websky simulation employs a different pressure profile
which is calibrated with the hydrodynamical simulations \citep{Battaglia2012b}
and the non-thermal pressure of this model is higher than our ICM model \citep{Battaglia2012a}.
Therefore, the difference appears in the tSZ power spectrum as the smaller amplitude.

\begin{figure}
  \includegraphics[width=\columnwidth]{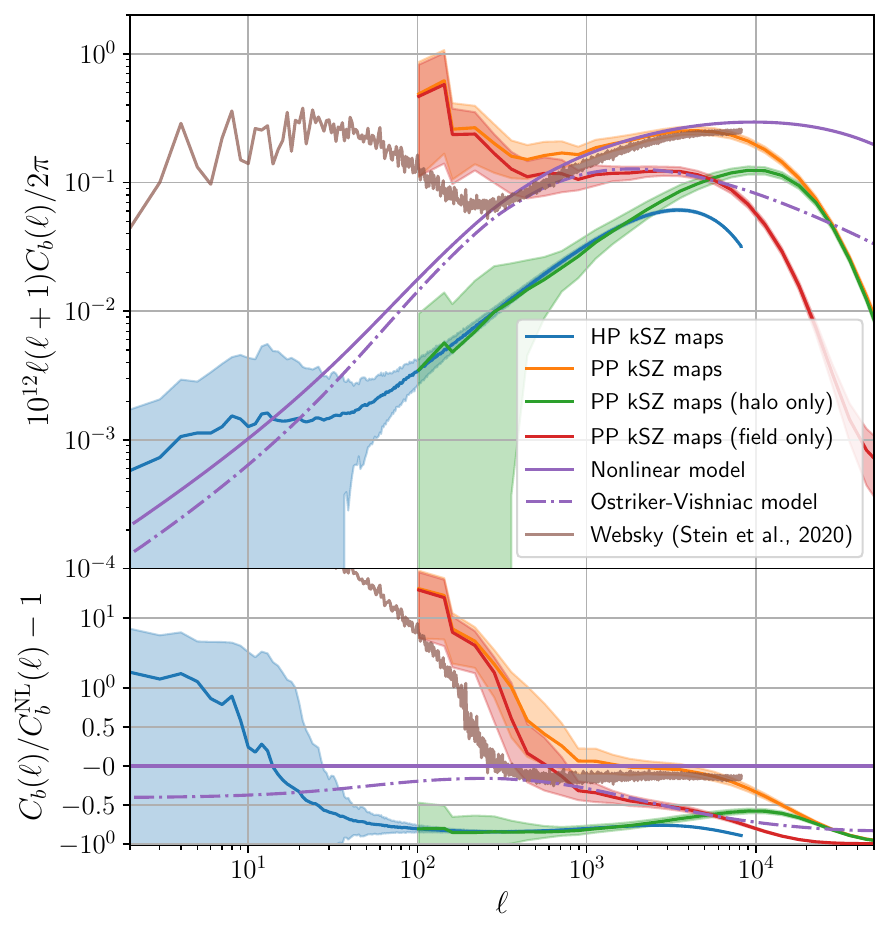}
    \caption{The angular power spectra of mock all-sky HP and flat-sky ($5 \times 5 \, \mathrm{deg}^2$) PP kSZ maps.
    The solid lines and the shaded regions correspond to
    the mean and the standard deviation of 108 HP and PP maps, respectively.
    The same caveat about the magnitude of the errors noted in Figure~\ref{fig:cl_tSZ} also holds here.
    For comparison, analytical results with the nonlinear model and Ostriker--Vishniac model and Websky mock all-sky simulation \citep{Stein2020} are also shown.
    For PP kSZ maps, we also show results with only halo and field particles.}
    \label{fig:cl_kSZ}
\end{figure}

Figure~\ref{fig:cl_kSZ} shows power spectra of kSZ HP and PP maps along with theoretical predictions
and the Websky mock all-sky map result \citep{Stein2020}.
For the kSZ power spectra, the field component dominates the halo component
at all scales.
For the halo components, there is a good agreement between the power spectra from the HP and PP maps.
At scales of $10^2 \lesssim \ell \lesssim 10^3$,
the result of PP maps is larger than the theoretical prediction for two reasons.
First, the contribution from the longitudinal momentum field becomes dominant on large scales \citep{Park2016}.
Secondly, our light-cone output for PP maps consists of discrete snapshots;
the discontinuity between snapshots may hamper the mode cancellation in the line-of-sight direction,
and there is no time evolution within each snapshot.
As a result, the power on large scales may be large
(see Appendix~\ref{sec:light_cone} for more detailed discussions). At scales of $10^3 \lesssim \ell \lesssim 10^4$,
the results of PP maps agree with the theoretical prediction based on the nonlinear model at a better than $10\%$ level.
The OV power spectrum is much smaller than the nonlinear power spectrum,
demonstrating that the nonlinear effect at small scales propagates to the kSZ power spectrum on large scales
due to the mode coupling in the momentum power spectrum (Eq.~\ref{eq:perp_power}).
The OV power spectrum is consistent with the PP field-only maps
at scales of $10^3 \lesssim \ell \lesssim 10^4$, demonstrating that the linear theory describes the large-scale kSZ power well. 
At scales of $\ell \gtrsim 10^4$, all simulation results show
sudden damping caused by smoothing due to the projected SPH kernel.
This feature suggests that our current simulation resolution cannot resolve the relevant small-scale structures
and require higher mass resolution to improve the level of agreement with the theoretical prediction at these scales.
However, the scales of $\ell \simeq 10^4$ are already small enough scales accessible
to many of the ongoing and upcoming kSZ observations in the coming decade,
so our current resolution should be sufficient for these surveys.
As for the comparison for the Websky mock simulation,\footnote{We measured the power spectrum
from the Websky late-time kSZ temperature fluctuation map
projected up to the redshift $z = 4.5$, excluding the contribution from patchy reionization.}
our PP maps are in good agreement on the scales of $10^3 \lesssim \ell \lesssim 10^4$.
Similarly to the tSZ power spectrum, we scale the power spectrum
with the scaling relation of $C_b \propto \sigma_8^{4.46} \Omega_\mathrm{b}^{2.12} h^{1.65}$
\citep{Shaw2012}.\footnote{This scaling relation is derived from the power spectrum at multipole $\ell = 3000$
but we apply this scaling at the whole multipole range
since the scaling relation varies only weakly with respect to the multipoles.
See Table~3 of \citet{Shaw2012} for the scaling relation at different multipoles.}
Therefore, the apparent offset in the kSZ power spectrum's amplitude likely originates from the different ICM model.
In addition, the scale where the longitudinal mode becomes dominant is larger in the Websky map than in our PP maps.
Since the Websky map is constructed using the continuous light-cone,
the incomplete cancellation of longitudinal modes occurs only at the sharp cutoff at the edge of the light-cone.

\subsection{Computational Speed and Efficiency}
Our HP and PP algorithms are computationally efficient. For all-sky HP maps,
there are $4.1 \times 10^7$ halos in total, and it took $170$ minutes ($907$ core hours) to produce a single map
with 320 cores: 8 nodes on 2.4 GHz Intel Xeon Skylake 6148 with 40 cores per node.
For light-cone HP and PP maps, there are $5.6 \times 10^5$ halos
and it took $1.5$ and $62$ minutes ($5.6$ and $231$ core hours)
for the production of a single HP and PP map, respectively.
In these calculations, we used 224 cores: 2 nodes on
2.7 GHz Intel Xeon Platinum 8280 with 112 cores per node.
HP is faster than PP since PP also calculates contributions from the diffuse component.
Nevertheless, these algorithms are much more computationally efficient than running
a full physics hydrodynamical simulation.

\section{Conclusions}
\label{sec:conclusions}
In this paper, we presented two approaches to produce mock tSZ and kSZ maps from dark matter
only $N$-body simulations: halo-based pasting (HP) and particle-based pasting (PP).
The HP method employs the halo model prescription and captures a bulk of the tSZ effect,
which are sourced primarily by hot gas embedded within dark matter halos,
but neglects the contributions from other parts of the cosmic web structures, such as filaments.
However, for the kSZ effect, an appreciable fraction of the signal comes from diffuse gas,
which makes the halo model prescription insufficient for modelling the kSZ effects.
Furthermore, halos are assumed to be spherically symmetric,
which is not always the case in the real Universe and leads to the irregular shape of SZ signals.
To address these issues, we developed an alternative pasting algorithm: the particle-based pasting (PP) method.
This method utilises particle distribution in $N$-body simulations
and incorporates diffuse components outside of halos and asphericity of halos.

In both methods, we make use of the ICM model \citep{Ostriker2005,Bode2007,Bode2009,Shaw2010},
where gas distribution is solved with analytic treatments.
In this ICM model, there are free parameters that regulate feedback efficiency,
non-thermal pressure, stellar-to-halo-mass relation, etc.,
and these parameters have been calibrated using X-ray observations of SZ-selected galaxy clusters \citep{Flender2017}.

To validate our map-making algorithms,
we compared power spectra measured from mock SZ maps based on HP and PP methods to theoretical predictions.
For the tSZ effect, the power spectra of HP maps are consistent with the halo model prediction within $10\%$ level since both assume halos' sphericity.
The amplitude of the power spectra of the HP maps is lower than the halo model prediction
at scales of $\ell \gtrsim 1000$ due to lack of resolution.
On the other hand, the power spectra of PP maps are smaller by $20\%$
compared with the halo model prediction and HP maps.
The suppression of the power spectrum is caused by the asphericity of halos,
which are neglected in the halo model, and the resultant amplitude becomes closer
to the \textit{Planck} 2015 measurement \citep{Bolliet2018}.
Thus, the incorporation of the halo asphericity has the potential
to resolve the discrepancy of amplitudes between the halo model and measurements.
For the kSZ effect, the results are markedly different between HP and PP kSZ maps.
The dominant source of the kSZ signal is the diffuse gas component, which is only included in PP maps.
The amplitude of power spectra for PP maps is 10 times larger than HP maps,
in which only gas within the virialized regions of dark matter halos is considered.
Suppose we restrict the particles within halos for the PP algorithm. In that case,
the power spectra are consistent for scales of $\ell \lesssim 1000$ within the $10\%$ level,
where the HP methods fail to capture the relevant small-scale gas physics within halos.
Compared with the theoretical prediction, our PP maps give a good match at scales of $500 \lesssim \ell \lesssim 10^4$. Still, there are discrepancies
at large ($\ell \lesssim 500$) and small ($\ell \gtrsim 10^4$) scales
due to insufficient cancellation of parallel mode and resolution, respectively.

Our mock simulation suite has broad applications for modeling and interpreting data from ongoing and future SZ surveys.
First, our method requires only halo catalogues for HP and particle distribution for PP.
Thus, we do not have to run computationally expensive hydrodynamical simulations.
We demonstrated that map production takes about 3 hours for all-sky HP maps and
1 hour for $5 \times 5 \, \mathrm{deg}^2$ PP maps,
which enables fast production of a large number $\mathcal{O}(10^3)$ of independent mock SZ maps.
The baryon pasted mock maps will be useful for (a) estimating the covariance matrix of SZ statistics
as well as (b) testing observational analysis pipelines
by incorporating various observational systematics (e.g., survey mask, detector noise, and selection function).

Although this paper has focused on modelling SZ observables,
our methodology can be extended to model other observables sourced by gas or dark matter,
such as X-ray emission, dispersion measure probed by fast radio bursts,
weak lensing, and cosmic infrared background.
We will address the validation of the baryon pasting algorithms presented
in this paper with hydrodynamical simulations, e.g., IllustrisTNG \citep{Nelson2019},
in a subsequent paper.
In addition, the cross-correlations between these observables have also been measured in observations,
and our pasting methods are powerful approaches
in applications to cross-correlations.
In future work, we plan to make use of on-the-fly continuous light-cone outputs
from \texttt{Gadget-4} \citep{Springel2021} for efficient production of various observables
and the largest $N$-body simulations such as Uchuu simulations \citep{Ishiyama2021},
which enables a plethora of multi-wavelength cross-correlation studies.

\section*{Acknowledgements}
The authors acknowledge Erwin Lau and Srinivasan Raghunathan for their useful comments on the manuscript.
KO acknowledges Joe DeRose, Zack Li, and Jia Liu for useful discussions.
KO was supported by JSPS Research Fellowships for Young Scientists.
This work was supported by Grant-in-Aid for JSPS Fellows Grant Number JP21J00011 and Yale University.
Numerical simulations were carried out on Cray XC50 at the Center for Computational Astrophysics in
National Astronomical Observatory of Japan and Yukawa-21 at Yukawa Institute for Theoretical Physics at Kyoto University.
The authors acknowledge Yale Center for Research Computing for data sharing support.

\section*{Data Availability}
The 108 all-sky and flat-sky tSZ and kSZ maps based on HP and PP algorithms are publicly available
and the details are found in \url{https://github.com/0satoken/BP_SZ_maps}.



\bibliographystyle{mnras}
\bibliography{main}



\appendix

\section{Effect of discreteness of the light-cone}
\label{sec:light_cone}
In order to investigate how the truncation due to discrete snapshots
in the light-cone output (Figure~\ref{fig:light_cone})
affects the kSZ effect, we employ continuous light-cone output with \texttt{Gadget-4}.
We have run a simulation with $N = 512^3$ particles and
the box size on a side $L = 1 \, h^{-1} \, \mathrm{Gpc}$.
We enabled light-cone output with all-sky coverage up to the redshift $z = 3.741$,
where particles are output when they
move across the backward light-cone (see Section~7.5 of \citet{Springel2021} for details).
The simulation box is replicated 720 times at maximum to cover the volume
up to $z = 3.741$ under periodic boundary conditions.

We construct an all-sky kSZ map from the light-cone output.
Since we are primarily interested in large-scale features in this exercise,
we did not distinguish halo and field particles in the light-cone output.
Instead, we treat all particles as field particles,
and the contribution of the kSZ signal from each particle is given by Eq.~\eqref{eq:field_b}.
The map is pixellated using \texttt{healpix} with $N_\mathrm{side} = 1024$.
Because the \texttt{healpix} pixel grid is not regular,
we cannot use integrated SPH kernels (Eq.~\ref{eq:integrated_kernel}).
Instead, the contribution is assigned to the nearest pixel.

\begin{figure}
  \includegraphics[width=\columnwidth]{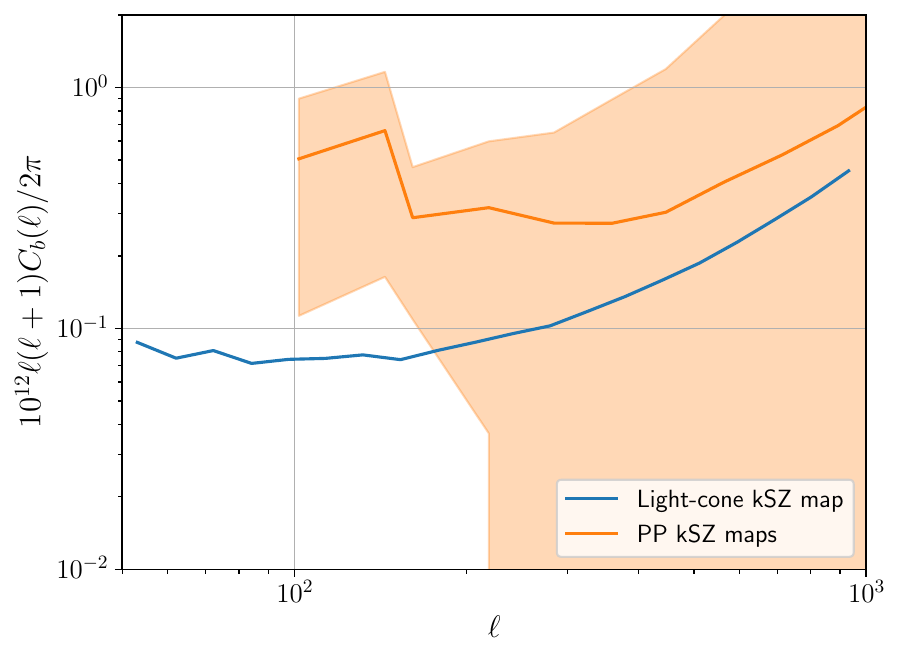}
    \caption{Power spectra of the all-sky light-cone kSZ map.
    For comparison, the results of PP kSZ maps without the ICM model are also shown,
    and the solid line and the shaded region correspond to
    the mean and the standard deviation of 108 PP kSZ maps, respectively.}
    \label{fig:cl_lightcone_kSZ}
\end{figure}

Figure~\ref{fig:cl_lightcone_kSZ} shows the kSZ power spectra of the all-sky light-cone map and PP kSZ maps.
Although the shot noise dominates at small scales,
the excess power appears at large scales ($\ell \lesssim 300$) in the PP maps,
because of incomplete cancellation of longitudinal mode.
The power spectrum of the continuous light-cone map, on the other hand, is relatively flat,
since the smoothness of outputs ensures the cancellation and suppresses the power at large scales.

\bsp 
\label{lastpage}
\end{document}